%% file: arxivversion.tex
\documentclass[graybox]{svmult}

\usepackage{mathptmx}       % selects Times Roman as basic font
\usepackage{helvet}         % selects Helvetica as sans-serif font
\usepackage{courier}        % selects Courier as typewriter font
\usepackage{type1cm}        % activate if the above 3 fonts are
\usepackage{makeidx}         % allows index generation
\usepackage{graphicx}        % standard LaTeX graphics tool
\usepackage{multicol}        % used for the two-column index
\usepackage[bottom]{footmisc}% places footnotes at page bottom
\usepackage{color}
\usepackage[dvipsnames]{xcolor}
\usepackage{ifpdf}

\catcode`\@=11
\newcommand{\gapprox}{\mathrel{\mathpalette\@versim>}}
\newcommand{\lapprox}{\mathrel{\mathpalette\@versim<}}
\newcommand{\propapprox}{\mathrel{\mathpalette\@versim\propto}}
\newcommand{\@versim}[2]
  {\lower3.1truept\vbox{\baselineskip0pt\lineskip0.5truept
\ialign{$\m@th#1\hfil##\hfil$\crcr#2\crcr\sim\crcr}}}
\catcode`\@=12
\newcommand{\cb}[1]{ {\colorbox{yellow}{#1} } }

\begin{document}

\title*{Dynamical Evolution and Radiative Processes of Supernova Remnants}
% Use \titlerunning{Short Title} for an abbreviated version of
% your contribution title if the original one is too long
\author{Stephen P. Reynolds}
% Use \authorrunning{Short Title} for an abbreviated version of
% your contribution title if the original one is too long
\institute{North Carolina State University, Raleigh, NC 27695-8202, USA,
\email{reynolds@ncsu.edu}}
%
% Use the package "url.sty" to avoid
% problems with special characters
% used in your e-mail or web address
%
\maketitle

\abstract{I outline the dynamical evolution of the shell remnants of
  supernovae (SNRs), from initial interaction of supernova ejecta with
  circumstellar material (CSM) through to the final dissolution of the
  remnant into the interstellar medium (ISM).  Supernova ejecta drive
  a blast wave through any CSM from the progenitor system; as material
  is swept up, a reverse shock forms in the ejecta, reheating them.
  This ejecta-driven phase lasts until ten or more times the ejected
  mass is swept up, and the remnant approaches the Sedov or
  self-similar evolutionary phase.  The evolution to this time is
  approximately adiabatic.  Eventually, as the blast wave slows, the
  remnant age approaches the cooling time for immediate post-shock
  gas, and the shock becomes radiative and highly compressive.
  Eventually the shock speed drops below the local ISM sound speed and
  the remnant dissipates.  I then review the various processes by
  which remnants radiate.  At early times, during the adiabatic
  phases, thermal X-rays and nonthermal radio, X-ray, and gamma-ray
  emission dominate, while optical emission is faint and confined to a
  few strong lines of hydrogen and perhaps helium.  Once the shock is
  radiative, prominent optical and infrared emission is produced.
  Young remnants are profoundly affected by interaction with often
  anisotropic CSM, while even mature remnants can still show evidence
  of ejecta.}

\abstract*{}

\section{Introduction}
\label{sec:1}

In this review, I shall first give a brief overview of the dynamical
evolution and radiative properties of SNRs.  I then provide a more
detailed discussion of each. I shall assume a basic familiarity with
fluid dynamics, shock waves, and radiative processes, at the level of
Shu (1991) and Rybicki \& Lightman (1979).  General physics of the
interstellar medium is covered in Spitzer (1978) and Draine (2011).
Subsequent chapters in this Section cover in more detail most of
the issues raised in this review.

\subsection{Evolutionary Overview}
\label{subsec:1}

As described in previous chapters, stellar ejecta are accelerated by
the emerging shock wave to speeds ranging as high as 30,000 km
s$^{-1}$, but with average values of order 5,000 km s$^{-1}$ for
core-collapse (CC) explosions and 10,000 km s$^{-1}$ for Type Ia
events.  This material may be quite anisotropic, and it initially
encounters material which may have been substantially modified by the
progenitor system.  This {\colorbox{yellow} {circumstellar material
    (CSM)}}\index{circumstellar material (CSM)} is likely also to be
quite anisotropic, most likely resulting from a stellar wind which may
have an azimuthal density dependence, or from interaction of the
progenitor star with a binary companion.  For Type Ia events, it is
also possible that the immediate SN environment is almost devoid of
material or containing only typical ISM.

In either case, the SN {\colorbox{yellow} {blast wave}}\index{blast wave}
or {\colorbox{yellow}{forward shock}}\index{forward shock} 
begins to decelerate almost immediately as it moves into this
surrounding CSM or ISM, heating it to X-ray emitting temperatures,
with a ``contact discontinuity,'' across which the pressure is roughly
constant, separating shocked CSM/ISM from ejecta.  The rapid expansion
at early stages cools the ejecta adiabatically to very low
temperatures, so that even a small amount of deceleration of the blast
wave results in a velocity difference that is greater than the sound
speed in the cold ejecta, and a {\colorbox{yellow} {``reverse shock''}}
\index{reverse shock} is born, facing inward, and reheating the
ejecta.  In even the youngest known SNRs, this reverse shock is
inferred to be present.

\begin{figure}[b]
%\sidecaption
%\centerline{\includegraphics[scale=0.4]{shocks.eps}}
\centerline{\includegraphics[scale=0.4]{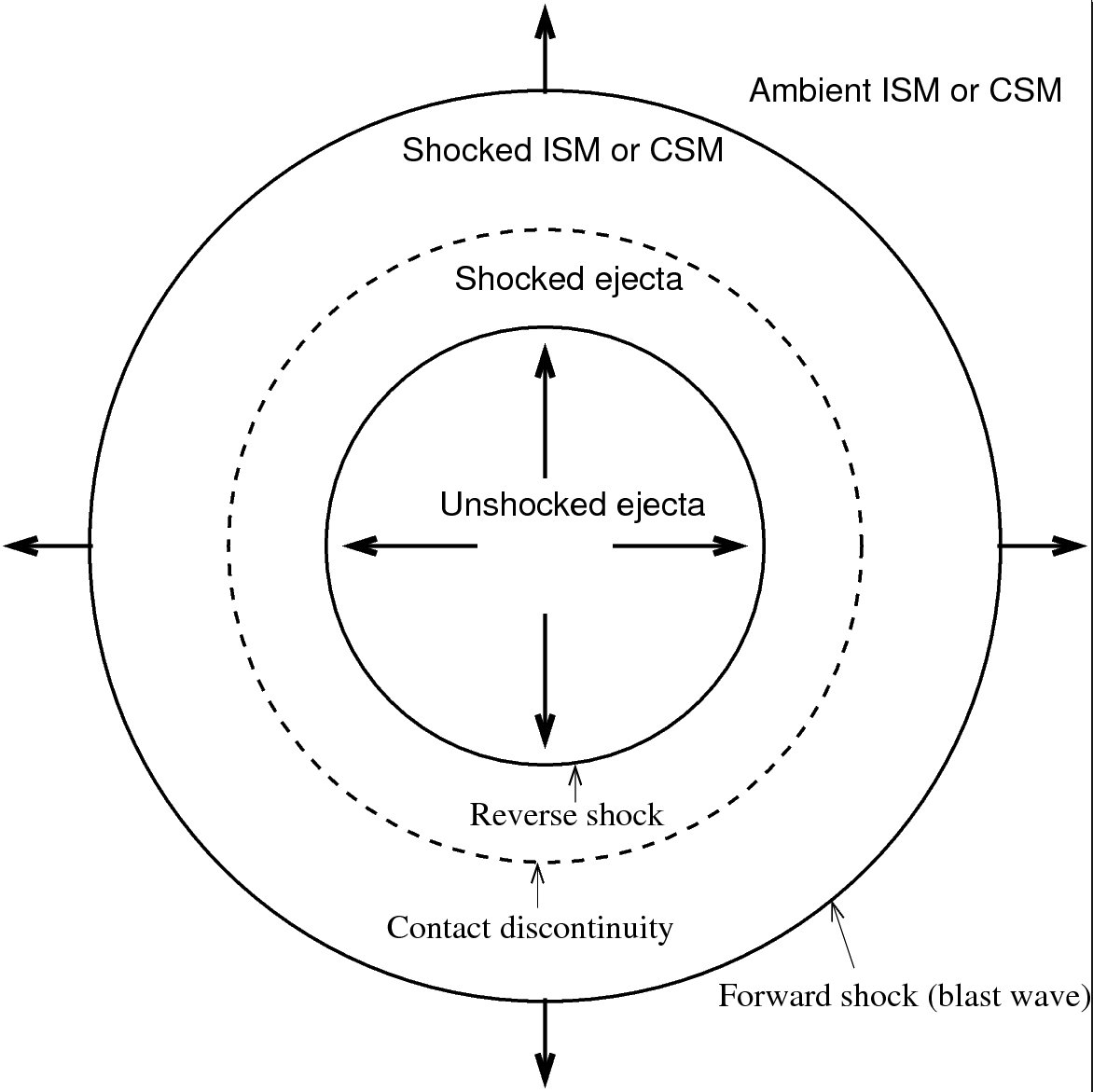}}
\caption{Schematic of the two-shock structure of a SNR in the
ejecta-driven stage.  Rapidly moving, cold unshocked ejecta are
heated and decelerated at the reverse shock.  Hot ejecta are separated
from shocked ambient material at a contact discontinuity.  The
forward shock or blast wave heats and accelerates ambient ISM
or CSM.}
\label{fig:shocks}
\end{figure}

The evolutionary stage in which both forward and reverse shocks are
present can last for hundreds to thousands of years.  It is sometimes
called the {\colorbox{yellow}{``ejecta-driven''}}\index{ejecta-driven}
stage.  During this stage, the remnant evolution depends on the
density structure in the ejecta as well as in the surrounding
material.  Observational signatures of this phase typically center on
the identifiable presence of enhanced elemental abundances in X-ray
spectra.  Relative contributions from the forward and reverse shocks
depend on density structure as well.  However, the energy radiated is
a small fraction of the kinetic energy released in the explosion, so
this evolution is approximately adiabatic.  The progressive
deceleration of the blast wave can be conveniently described with an
{\colorbox{yellow}{``expansion parameter''}}\index{expansion
  parameter} $m$ defined by $R_s \propto t^m$, with $R_s$ the
(forward) shock radius.  Undecelerated expansion with $m = 1$ almost
immediately gives way to $m < 1$, and various analytic solutions exist
describing subsequent evolution.  However, numerical simulations
demonstrate the gradual decrease in $m$ as swept-up material comes to
dominate the expansion.

For constant-density ambient material, after about ten times the
ejected mass has been swept up, the value of $m$ approaches 0.4, its
value for the idealized {\colorbox{yellow}{Sedov self-similar
    solution}}\index{Sedov solution} for a point explosion in a uniform
  medium.  (There is also a Sedov solution for expansion into a
  power-law density gradient $\rho \propto r^{-s}$; for $s = 2$,
  appropriate for a steady spherically symmetric stellar wind, $m$
  approaches $2/3.$ However, this situation may not often be realized
  in practice.) Thus both the ejecta-driven and Sedov phases can be
  termed adiabatic.  (The ejecta-driven phase is still occasionally
  referred to as the ``free-expansion'' phase, but this is not really
  accurate.)

For CC remnants, a neutron star is likely to be present.  If it
functions as a pulsar, it can inflate a bubble of relativistic
particles and magnetic field, a pulsar-wind nebula (PWN), in the
remnant interior.  For a young, luminous pulsar, the PWN can
expand and overtake inner ejecta, driving a shock into them with
possible observational consequences.  However, for all but 
exceptional cases, the pulsar energy input is not sufficient to
alter the gross evolution of the shell SNR.  PWNe are the subject
of a later chapter.  

\begin{figure}[b]
%\centerline{\includegraphics[scale=0.35]{g1.9yr11b.eps}
%  \hskip0.1truein \includegraphics[scale=0.32]{casaradio.eps}
% \hskip0.05truein \includegraphics[scale=0.25]{casa3bandsc.eps}
\centerline{\includegraphics[scale=0.35]{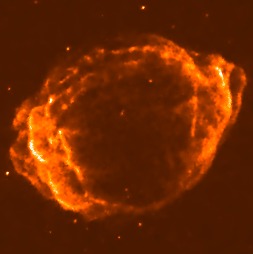}
  \hskip0.1truein \includegraphics[scale=0.32]{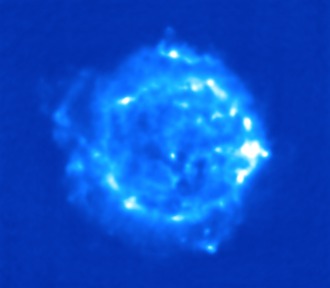}
 \hskip0.05truein \includegraphics[scale=0.25]{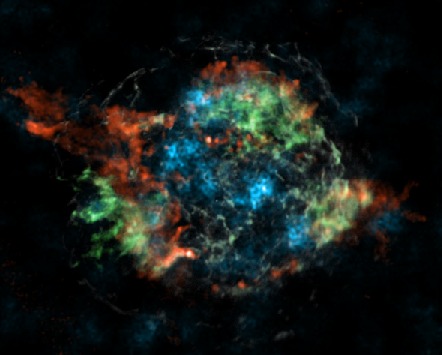}
 }
\caption{Remnants of the two most recent known supernovae in the Milky
  Way.  Left: {\colorbox{yellow}{G1.9+0.3}}\index{G1.9+0.3} (about
  1900 CE) (X-rays) (K.~Borkowski).  Center: \cb{Cassiopeia
    A}\index{Cassiopeia A} (5 GHz, VLA; DeLaney et al.~2014).  Right:
  Cassiopeia A (about 1680 CE).  Green: Si band.  Red: Fe K$\alpha$
  band.  (both with {\sl Chandra}; U.~Hwang).  Blue: $^{44}$Ti
  emission (68 keV) with NuSTAR (Grefenstette et al.~2014).}
\label{fig:g1casa}
\end{figure}

As the blast wave decelerates, eventually the timescale for radiative
cooling of the shocked material becomes comparable to the remnant age.
(Cooling is typically by UV, optical, and near-IR fine-structure
transitions of astrophysically common elements such as C, O, and Fe.)
This is normally for shock speeds $v_s \sim 200$ km s$^{-1}$, only
weakly dependent on density, with corresponding ages of order 10,000
or more years.  Once cooling is important, deceleration is more rapid,
though the continuing presence of hot gas in the SNR interior, where
cooling times are longer, continues to operate in what is called a
``pressure-driven snowplow,'' with $m \sim 0.3.$ If that pressure is
negligible, material essentially coasts, conserving (local) momentum,
with $m \rightarrow 0.25.$ However, by these late stages, most
remnants have been interacting with inhomogeneous ISM for some time,
and are quite irregular, with properties varying substantially with
position in the remnant.  Densities in cooling shocks can be quite
high, as the compression may be limited only by magnetic pressure, so
radiative-phase remnants can be quite bright in optical emission.
Eventually, shock speeds become comparable to local sound speeds and
the SNR dissipates into the ISM.

\subsection{Radiation Overview}
\label{subsec:2}

At different stages, SNR radiation is dominated by different
processes.  After the initial SN light has declined, the rapid
expansion of the ejecta cools them to very low temperatures.  However,
the very strong, highly supersonic blast wave heats surrounding
material to X-ray-emitting temperatures.  Since all the relevant
astrophysical shock waves in SNRs are collisionless (gas is heated not
by binary collisions among particles but by interaction with a
magnetic field), the particle distribution downstream is not perfectly
Maxwellian.  Instead, {\colorbox{yellow}{diffusive shock acceleration
    (DSA)}}\index{ diffusive shock acceleration} of a small fraction
of electrons crossing the shock produces a (nearly) power-law
nonthermal tail attached to the thermal peak of the electron energy
distribution.  The tail normally extends to relativistic energies,
where electrons can radiate {\colorbox{yellow}{synchrotron radiation}}
\index{synchrotron radiation}
at radio
wavelengths.  For appropriate conditions, the synchrotron component
can extend all the way to the X-ray band.  Optical emission at early
stages is faint.  These fast shocks are called 
{\colorbox{yellow}{``nonradiative;''}}\index{nonradiative}
cooling times of the shocked gas are initially much longer than the
shock age, so relatively little radiation is produced, and the
compression ratio $r$ has the value appropriate for a strong (highly
supersonic) adiabatic shock into monatomic gas, $r = 4$.  However,
some radiation can be detected from such shocks; if the remnant
expands into neutral material, hydrogen can be excited before being
ionized and radiate Lyman and Balmer-series photons. Infrared
radiation at early stages is predominantly thermal radiation from 
{\colorbox{yellow}{dust}}\index{dust}
grains heated by collisions with hot gas; the temperature of that
infrared emission is a good diagnostic of plasma density.

As the blast wave decelerates, the initially weak and radiative
reverse shock strengthens and becomes non-radiative. The reverse shock
re-heats the ejecta that overtake it, rendering them observable in
X-ray emission.  SNRs in this stage primarily radiate radio
synchrotron emission and thermal X-ray emission from roughly
solar-abundance gas behind the blast wave and from enhanced abundances
of heavy elements behind the reverse shock.  Disentangling these two
contributions is a significant challenge in studying ejecta-driven
SNRs.

The process of collisional ionization of heavier elements in either
the shocked CSM/ISM or the shocked ejecta is not instantaneous.  In
fact, plasmas in young SNRs are typically underionized, that is, at a
lower stage of ionizaton than would be the case for a gas in
equilibrium at the observed temperature.  This \cb{non-equilibrium}
\cb{ionization (NEI),}\index{non-equilibrium ionization} as it is
called, means that X-ray spectra depend on a parameter $\tau \equiv
\int n_e\, dt$, the \cb{``ionization timescale,''}\index{ionization
  timescale} which controls the degree of ionization of the plasma.
In a shock wave, gas with all values of $\tau$ from zero up to the
shock age can emit radiation.  When $\tau$ exceeds a few times
$10^{12}$ cm$^{-3}$ s, plasma is close to collisional ionization
equilibrium (CIE).  For typical ambient densities of order 1
cm$^{-3}$, this occurs in $\sim 30,000$ yr. The ionization state of
the plasma at this stage does not affect the overall dynamics,
however.

\begin{figure}[b]
\includegraphics[scale=0.28]{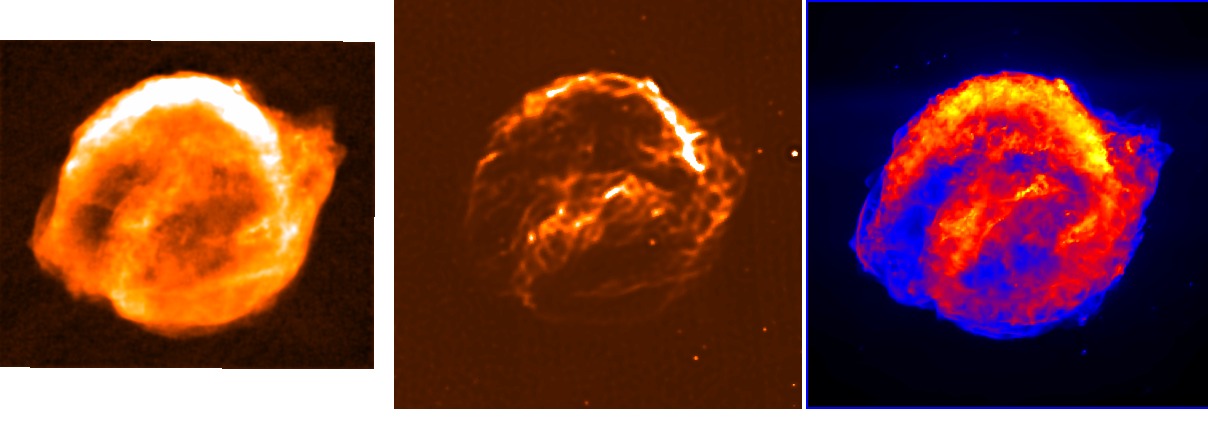}
\caption{The remnant of \cb{Kepler's supernova of 1604
    CE}\index{Kepler}.  Left: radio (VLA at 5 GHz; T.~DeLaney).
  Center: {\sl Spitzer} MIPS at 24 $\mu$m (deconvolved; K.~Borkowski).
  Right: {\sl Chandra} between 0.3 and 7 keV (Reynolds et al.~2007).}
\label{fig:Kepler3}
\end{figure}

\begin{figure}[b]
\includegraphics[scale=0.25]{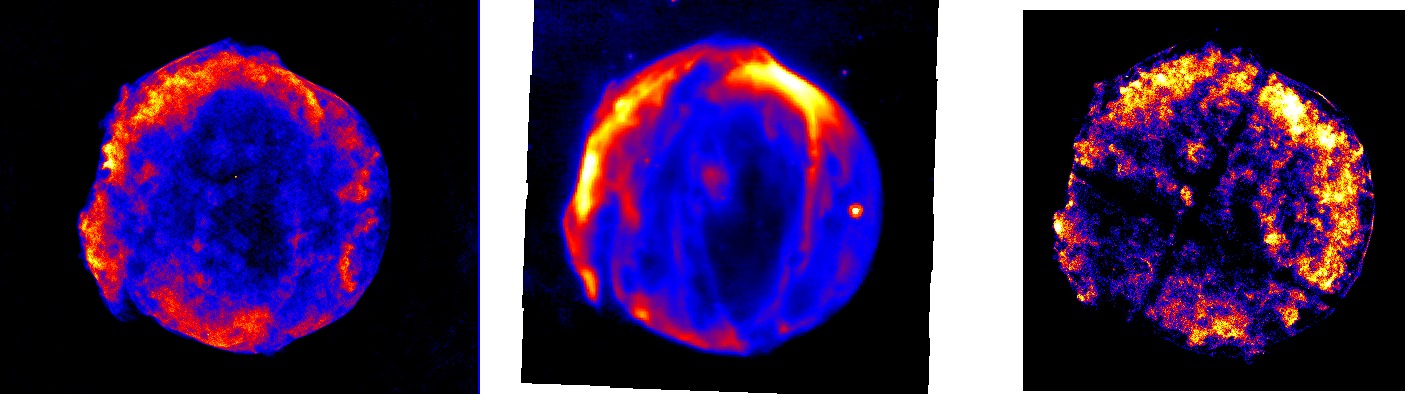}
\caption{The remnant of \cb{Tycho's supernova of 1572 CE}\index{Tycho}.  Left:
radio (VLA at 5 GHz; Reynoso et al.~1997).  Center:  {\sl Spitzer} MIPS
at 24 $\mu$m (Williams et al.~2013).  Right: {\sl Chandra} image (NASA/CXC).}
\label{fig:Tycho3}
\end{figure}

Infrared emission can be produced by radiation from collisionally
heated grains in either the shocked CSM/ISM or, if dust is formed in
the cool ejecta, in the post-reverse-shock region.  Line emission can
be detected from unshocked ejecta in regions of particularly high
density.  Finally, a few of the youngest remnants produce detectable
emission that is not related to shocks at all, but results from the
decay of radioactive $^{44}$Ti into $^{44}$Sc and then $^{44}$Ca, with
the emission of hard X-ray and gamma-ray nuclear de-excitation lines
and a line at 4.1 keV from filling the vacancy resulting from the
electron capture decay of $^{44}$Ti to $^{44}$Sc.

Once the reverse shock has disappeared and the remnant is fully in the
Sedov stage, spectral signatures of enhanced-abundance ejecta may
still be present in X-rays in the interior.  However, the shocked
CSM/ISM mass dominates the shocked ejecta and the integrated spectrum.
IR continuum from heated grains can still be produced.

The onset of radiative cooling dramatically alters the spectral-energy
distribution (SED) of a SNR.  Now, UV, optical and IR permitted,
forbidden, and fine-structure transitions produce optically bright
spectra dominated in optical by low ionization stages of elements like
sulfur and oxygen.  In fact, for SNRs in external galaxies, a powerful
method of identifying radiative-stage remnants is the ratio of [S II]
$\lambda\lambda$ 6717, 6731 to H$\alpha$ flux, which is very much
different for a \cb{radiative shock}\index{radiative shock} than for
photoionized H II regions.  Since adiabatic-phase SNRs do not produce
bright optical emission, this method does not identify them.  However,
most of the observable life of a SNR is spent in the later phases, so
a relatively small fraction of remnants in other galaxies is
overlooked.  For shock velocities below 200 km/s or so, X-ray emission
is now quite weak.  Nonthermal radio synchrotron emission from
electrons with GeV energies can persist and remains the most easily
observed observational signature of SNRs.  The high compression ratios
characteristic of radiative shock waves will also enhance the
synchrotron brightness.

Nonthermal emission at X-ray wavelengths and above can be observed in
a few remnants.  Sufficiently energetic relativistic electrons (and
positrons, if present) can produce not only synchrotron emission up to
X-rays, but \cb{inverse-Compton emission}\index{inverse-Compton
  emission} from upscattering any photon fields (cosmic microwave
background and possibly any locally strong IR or optical radiation)
and \cb{bremsstrahlung}\index{bremsstrahlung} from interaction with
thermal ions.  These leptonic contributions can extend to GeV and even
TeV photon energies.  In addition, the shock acceleration process is
expected to accelerate ions as well.  While they do not directly
radiate, they can inelastically scatter from thermal ions, producing
charged and neutral \cb{pions}\index{pions} (and secondary positrons and electrons).
The charged pions decay eventually to electrons and positrons, but the
neutral pions decay to pairs of gamma-rays once the
\cb{cosmic-ray}\index{ cosmic-ray} proton energies are sufficient to
produce $\pi^0$ particles (around 70 MeV).  This 
\cb{``hadronic'' process}\index{hadronic process}
is the only direct evidence for cosmic-ray ions in SNRs.  See Reynolds
(2008) for a review of supernova remnants with emphasis on high-energy
radiative processes.

\begin{figure}[b]
%\centerline{\includegraphics[scale=0.6]{sn1006xc.eps} 
%  \hskip0.1truein \includegraphics[scale=0.35]{g11.2x.eps}}
\centerline{\includegraphics[scale=0.6]{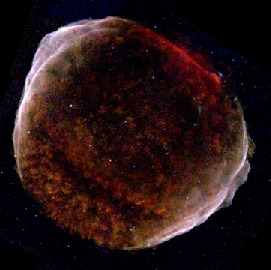} 
  \hskip0.1truein \includegraphics[scale=0.35]{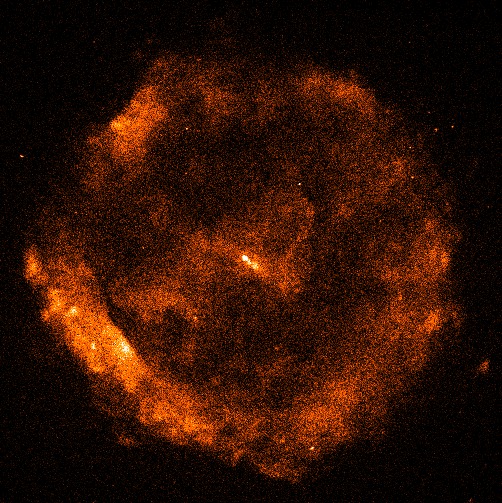}}
\caption{Left: Remnant of \cb{SN 1006 CE;}\index{SN 1006} (X-ray, {\sl
    Chandra} Winkler et al.~2014).  Right:
  \cb{G11.2--0.3}\index{G11.2--0.3}, roughly 2000 y old (X-ray, {\sl
    Chandra}; Borkowski et al.~2015).}
\label{fig:sn1006g11}
\end{figure}

\section{Dynamical Evolution}
\label{sec:2}

I shall now consider in more detail the dynamical evolution, dividing
the discussion into the phases outlined above:  Ejecta-driven,
Sedov, and Radiative.  

\subsection{Ejecta-Driven Evolution}
\label{subsec:3}

We can consider the ``initial conditions'' for supernova-remnant
evolution to be the distribution of ejected material once pressure
forces from the original explosion are negligible (``ballistic
expansion'').  This is certainly true for all but very exceptional
cases after a few weeks.  The density profile of expanding material is
determined by the density structure of the progenitor star and its
interaction with the shock wave that disrupts the star.  Early 1-D
hydrodynamic simulations showed that both CC (nondegenerate) and Type
Ia (degenerate) progenitors produced expanding profiles roughly
describable as a central region of roughly constant density and an
outer region of steeply declining density following an approximately
power-law density dependence, $\rho \propto r^{-n}$ with $n \sim 7$
for white dwarf progenitors and $n \sim 10 - 12$ for CC progenitors.
More extensive hydrodynamic simulations and analytic calculations have
refined these numbers somewhat. Dwarkadas \& Chevalier (1998) find
that exponential density profiles provide better fits to simulations
for Type Ia SNRs.  Matzner \& McKee (1999) used realistic 1D stellar
progenitor models for CC events and calculated the resulting ejecta
distribution after the passage of the original supernova shock wave.
The steep outer power-laws are reproduced, but there is a clear
density jump (corresponding to the original interface between the
progenitor's hydrogen envelope and interior) of about a factor of 3 --
10, within which the density is very roughly constant.

Of course, real supernovae are not likely to be perfectly spherical.
Rotation of the progenitor is an obvious cause of asymmetry, but in
addition, the fundamental explosion mechanism may be asymmetric.  If
the \cb{standing accretion shock instability}
\index{standing accretion shock instability} (SASI; Blondin et al. 2003)
or another convective instability is important for CC events, material
may be primarily ejected in one direction, with a high-velocity
neutron star moving off in the opposite direction.  In addition to
large-scale asymmetries such as these, there may be smaller-scale
clumping or other irregularities.  Ejecta clumps are seen ahead of the
average blast-wave radius in most young, and a few older, SNRs, and
high-density knots of fast-moving material with enhanced abundances
are seen in Cas A (roughly 330 yr old).  (See, for example, Winkler et
al.~2014 for SN 1006 and Hammell \& Fesen 2008 for Cas A.)  Radiative
knots in the interior of Kepler's SNR (CE 1604) may be CSM, but some
knots ahead of the blast wave are ejecta (Reynolds et al.~2007).  As
the $^{56}$Ni synthesized in a CC explosion decays, its decay products
will heat local material which will then expand into cooler ejecta
(the ``nickel bubble'' effect; Li et al.~1993).  The effect is
enhanced if the $^{56}$Ni is not uniformly distributed but in clumps
instead.  This effect may well produce inhomogeneities in SN ejecta
that are detectable in young SNRs.

\begin{figure}
%\centerline{\includegraphics[scale=0.35]{w28_3bands.eps} 
%  \hskip0.05truein \includegraphics[scale=0.3]{cygloophris.eps}
%  \hskip0.05truein \includegraphics[scale=0.4]{pupahris3.eps}}
\centerline{\includegraphics[scale=0.35]{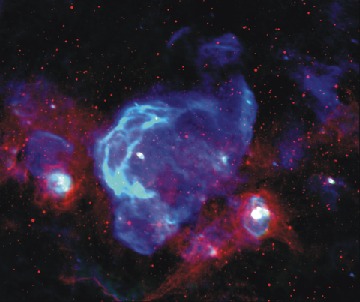} 
  \hskip0.05truein \includegraphics[scale=0.3]{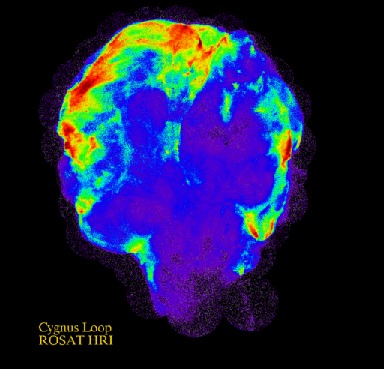}
  \hskip0.05truein \includegraphics[scale=0.4]{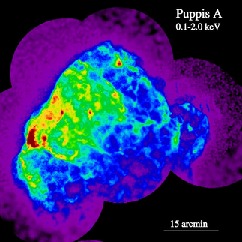}}
\caption{Three older SNRs.  Left:  Three-color image of \cb{W28.}
\index{W28}
Red: infrared; cyan: H$\alpha$; blue: radio (VLA) (NRAO/AUI/NSF; Brogan et al.)
Center:  \cb{Cygnus Loop}\index{Cygnus Loop} (X-ray; ROSAT) (NASA/GSFC).  Right:  \cb{Puppis A}\index{Puppis A}
(X-ray; ROSAT) (NASA/GSFC).}
\label{fig:3older}
\end{figure}

\begin{figure}
%\centerline{ \includegraphics[angle=270,scale=0.5]{s0shockr.ps}
%             \includegraphics[angle=270,scale=0.5]{s0profiler.ps}}
\centerline{ \includegraphics[angle=270,scale=0.5]{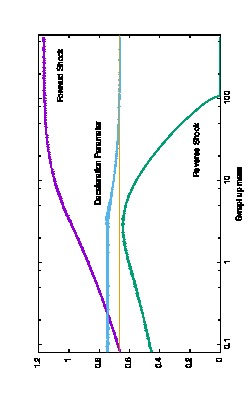}
             \includegraphics[angle=270,scale=0.5]{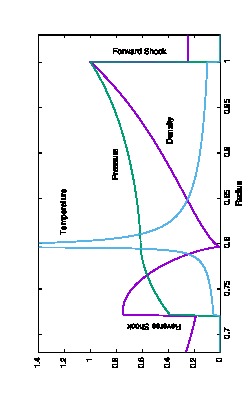}}
\caption{Left:  Shock locations and forward-shock expansion index $m$
for the case $n = 6$, $s = 0.$  Shock positions are scaled by $t^{0.4}$
(the Sedov value). Right:  Profiles of density, temperature,
and pressure during the self-similar phase for this calculation. (J.~Blondin,
private communication}
\label{fig:s0}
\end{figure}

\begin{figure}
%\centerline{ \includegraphics[angle=270,scale=0.5]{s2shockr.ps}
%             \includegraphics[angle=270,scale=0.5]{s2profiler.ps}}
\centerline{ \includegraphics[angle=270,scale=0.5]{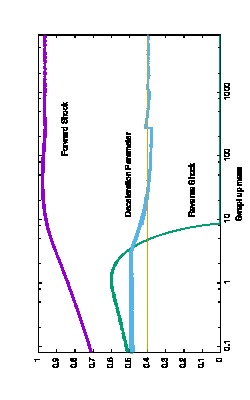}
             \includegraphics[angle=270,scale=0.5]{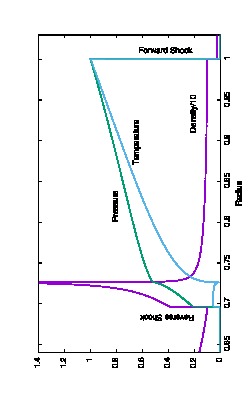}}
\caption{Left:  Shock locations and forward-shock expansion index $m$
for the case $n = 6$, $s = 2.$  Shock positions are scaled by $t^{0.4}$
(the Sedov value). Right:  Profiles of density, temperature,
and pressure during the self-similar phase for this calculation. (J.
Blondin, private communication}
\label{fig:s2}
\end{figure}

The immediate surroundings of the SN are almost certainly not uniform.
A steady-state constant-velocity wind in spherical symmetry produces a
$1/r^2$ density profile.  For spherical power-law supernova ejecta
with $\rho_e \propto r^{-n}$ encountering a surrounding medium with
$\rho \propto r^{-s}$, similarity solutions exist for the shock radii
and for the density and pressure profiles everywhere.  Outer shock
waves in such cases have values of the expansion index $m$
intermediate between 1 and the Sedov uniform-density value of 0.4: in
fact, $m = (n-3)/(n-s)$, as long as $n > 5$ and $s < 3$ (Chevalier
1982; Nadezhin 1985).  As required by the self-similarity, the ratio
between forward and reverse shock radii is constant; both move out,
but the reverse shock is overtaken by faster-moving ejecta.  The
character of the solutions is quite different depending on the outer
index.  For the steady wind value of $s = 2$, the density peaks at the
contact discontinuity; since the pressure is roughly uniform, the
temperature decreases there.  For uniform ISM ($s = 0$), the density
drops to zero at the contact discontinuity, and the temperature rises.
The same qualitative behavior occurs for decreasing, but
non-power-law, ejecta density profiles, such as the exponential
profile.  Figures~\ref{fig:s0} and \ref{fig:s2} 
show 1-D hydrodynamic simulations of a blast wave with
power-law density profile moving into a uniform medium and a steady
wind.  The shock radii are scaled by $t^{0.4}$ (the Sedov value).
Quantities are plotted as a function of the swept-up mass in units
of the ejected mass.

While the forward and reverse shocks in the ejecta-driven phase are
hydrodynamically stable, the region between them is not, in general.
The Rayleigh-Taylor instability of a heavy fluid supported by a light
one (or, in general, if the effective gravity ${\bf g}$ opposes the
density gradient $\nabla \rho$, ${\bf g} \cdot \nabla \rho < 0$)
operates as the deceleration provides an effective inward gravity, and
outer less dense material decelerates denser inner material.  The
growth rate of this instability is maximum at the contact
discontinuity.  It may produce turbulence that could accelerate
particles; this may explain the bright ring of radio emission seen
inside the outer blast wave of Cas A.

However, the surroundings may not even be spherically symmetric.  It
is becoming increasingly apparent that a large fraction of supernovae
occur in binary systems.  All SNe Ia, of course, result from binaries,
but several categories of CC event, such as SNe Ib/c and IIb, seem to
result from stripped cores which probably require a binary companion
(Smith et al.~2011).  The companion star may be near enough to decelerate
ejecta, possibly producing effects detectable in
remnants for hundreds of years.  Even if not, mass lost from either
the companion or the SN progenitor star itself is likely to be
asymmetric in the immediate neighborhood, probably focused into the
orbital plane in a disk wind.  (See Smith 2012 for a review of mass
loss in massive stars.)  A SN blast wave encountering an equatorially
enhanced CSM, with the whole system moving at high velocity, seems to
be the picture required to explain various features of Kepler's SNR
(Burkey et al.~2013).

Winds of supernova progenitor systems go through various phases.  A
massive OB star will have a fast wind while on the main sequence, but
as a red giant is likely to produce a slow, dense wind with a much
higher mass-loss rate.  Thus the CSM into which the SNR expands may be
highly structured.  The cumulative effect of the various wind phases
is generally to produce a low-density cavity or bubble (Castor,
McCray, \& Weaver 1975), eventually of roughly constant density except
near the star if mass-loss continues.  There is evidence that such
cavities can be produced by both CC and Type Ia progenitor systems;
the SN Ia remnant \cb{RCW 86}\index{RCW 86} is an excellent example of
  the latter (Williams et al.~2011).  A SN blast wave can race through
  the low-density cavity, remaining strong but not terribly luminous,
  until encountering the cavity wall, where a much slower transmitted
  shock moves into the wall while reflected shocks reheat the bubble
  interior.  Pre-SN wind phases, or for massive stars, episodic mass
  loss shortly before the SN, can result in a shell of CSM at a range
  of possible distances.  The SNR blast wave will slow on encountering
  the shell, but can accelerate again after traversing it.  This can
  result in overionized shocked plasma, with observational
  consequences (Yamaguchi et al.~2009).

Any structure in the ambient ISM will also affect SNR evolution.  SNe
Ia may be encountering such material after only a few hundred years;
there is evidence that Tycho's SNR (CE 1572) is interacting with such
material (Williams et al.~2013).  However, the ISM near Tycho appears
to have a substantial gradient in density, with densities a factor of
6 or more higher on one side than the other, in addition to
smaller-scale variations.  Hydrodynamic simulations of remnants
expanding into a smooth density gradient show that they can remain
remarkably round for hundreds or thousands of years, but that their
geometric centers can depart from the true explosion location by tens
of percent of the mean remnant radius, complicating any search for
remaining binary companions (Dohm-Palmer \& Jones 1996).

In general, while the simple spherically symmetric analytic pictures
are adequate for rough categorization of SNRs, detailed descriptions
of individual objects require hydrodynamical simulations, generally in
two and three dimensions.

\subsection{Sedov Evolution}
\label{subsec:4}

As long as the ejecta moving through the reverse shock are part of the
envelope material with a steep density profile, the ram pressure
upstream (inside) the reverse shock can keep it moving outward as
higher-density material arrives. However, when the roughly
constant-density central ejecta reach the reverse shock, this is no
longer the case, and the reverse shock will move back toward the
remnant center.  In 1-D analytic or numerical calculations, it
reflects strongly, but in 2 and 3D (and, presumably, in reality), the
reverse shock does not return exactly to the remnant center, but
reverberates in a complex way for a substantial transition period.
However, once all the ejecta have been shocked by the reverse shock,
we may assert that the remnant is fully in the Sedov stage.  Analytic
solutions (Sedov 1959) describe the run of density, pressure, and
temperature behind the shock.  The shock radius is given by a simple
anayltic expression: $R_s = 1.15 (E/\rho)^{1/5}t^{2/5}$, where $E$ is
the explosion energy and $\rho$ the ambient (uniform) density.  (We
presume a ratio of specific heats of 5/3.)

\begin{figure}
%\centerline{\includegraphics[angle=270,scale=0.5]{sedovs0r.ps}
%  \includegraphics[angle=270,scale=0.5]{sedovs2r.ps}}
\centerline{\includegraphics[angle=270,scale=0.5]{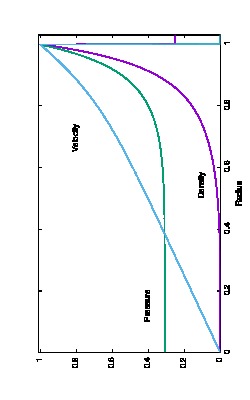}
  \includegraphics[angle=270,scale=0.5]{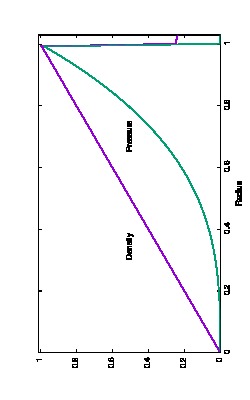}}
\caption{Left:  Sedov self-similar profiles of velocity, pressure,
and density for a blast wave into uniform-density surroundings.
Right:  Sedov profiles for a blast wave into an $r^{-2}$ density
profile.  Density and velocity profiles are identical in this
case; both are linear with radius.
(J.~Blondin, private communication). }
\label{fig:sedov}
\end{figure}

Analytic solutions have been produced by Truelove \& McKee (1999) that
describe the evolution in spherical symmetry from the early
ejecta-driven stage through the transition into Sedov evolution.
While there are self-similar solutions for a point explosion in a
medium with an arbitrary power-law density profile (Sedov 1959), in
almost all cases, uniform ambient density will be the best
approximation.  That produces a solution in which the post-shock
velocity is almost linear from zero at the center to 3/4 of the
blast-wave speed just behind the shock; the density drops steeply with
most of the mass within the outermost 10\% of the radius; and the
pressure drops slightly behind the shock, leveling out in most of the
interior at about 0.3 times its post-shock value.  This means the
temperature rises to unphysical values at small radii; in general, the
interior of a Sedov blast wave consists of very hot, low-density
material.

\subsection{Radiative Phase}
\label{subsec:5}

A perfectly spherical remnant in a perfectly uniform medium would
experience a sudden transition when its age reached a characteristic
cooling time for the gas (which depends on its composition; initial
cooling is from Fe, and, as the shock slows, from elements such as C
and O).  At that point, gas behind the shock would radiate away
significant amounts of energy and become much more compressible.  The
overall shock compression ratio would rise until some other mechanism,
perhaps magnetic fields, limited further compression.  Hydrodynamic
simulations in one and two dimensions show that the onset of cooling
is sudden, with the rapid formation of a cool dense shell subject (in
2D and 3D) to instabilities which rapidly disrupt it (Blondin et
al.~1998).  By this time, a typical remnant is so large that the ISM
it encounters is unlikely to be uniform.  Remnants may encounter
strong ISM inhomogeneities, such as dense molecular clouds, or may
produce ``blowouts'' into much less dense regions (e.g., 
\cb{3C 391;}\index{3C 391}
Reynolds \& Moffett 1993).  Dense material
near an SNR can serve as a target for escaping cosmic-ray ions, which
can produce gamma-rays from the decay of $\pi^0$'s produced in
inelastic collisions with thermal gas (e.g., W28: Aharonian et
al.~2008).

Since the remnant interior has a much lower density than the outer
regions, it has a much longer cooling time and that gas can remain
adiabatic long after the immediate post-shock gas has cooled.  It can
provide significant pressure to keep the expansion parameter
significantly above the value of 0.25 that characterizes purely
momentum-conserving evolution (Blondin et al.1998).  Remnants in this
stage are large, complex objects with typically large variations in
conditions at different locations. See Fig.~\ref{fig:3older} for
images of later-stage SNRs.

\section{Radiative Processes}
\label{sec:3}

As a remnant evolves through the stages outlined above, the
characteristic radiation it emits changes as well.  ``Prompt'' X-ray
and radio emission from the original supernova event, generally
attributed to interaction with a CSM of decreasing density, may take
years to decay away. At some point, thermal X-ray emission from the
increasing volume of shock-heated gas, and synchrotron emission from a
nonthermal electron distribution whose maximum energy rises with time,
produce true remnant X-ray and radio emission.  Only one Galactic
object has been caught in this rising phase: the remnant of the most
recent known Galactic supernova, G1.9+0.3, brightening at both radio
and X-ray wavelengths (Carlton et al.2011).  \cb{SNR 1987A}\index{SNR
  1987A} in the LMC is also brightening in both regimes (see the
review ``Supernova remnant from SN 1987A'' in this volume).

{\bf Ejecta emission.} The unshocked ejecta rapidly cool over the
months after the explosion to temperatures of order 100 K.  However,
the ejecta are illuminated by UV and soft X-ray emission from the
interaction region between the blast wave and newly formed reverse
shock.  This radiation can ionize elements with low ionization
potentials to produce near and mid-IR fine-structure lines from
species such as singly-ionized iron, argon, and neon.  In Cas A, where
this emission can be studied in detail, temperatures of a few thousand
K and densities $\lapprox 100$ cm$^{-3}$ are deduced in the unshocked
ejecta, although only in the denser regions; a considerable amount of
lower-density material could still be present.  (See Isensee et
al.~2010 for a thorough study.)

Shocked ejecta will typically be heated to X-ray-emitting
temperatures, $kT \gapprox 1$ keV.  However, evidence from comparing
X-ray and optical data indicates that the electron and ion
temperatures are not equal in fast shocks (Itoh 1978).  A shock
initially randomizes electron and ion speeds, so that the preshock
bulk velocity is (mostly) converted to postshock random velocity.  But
if ion and electron speeds are equal, proton and electron temperatures
would differ by the ratio of masses.  The timescale for electrons to
equilibrate in temperature with one another, $t_{ee}$, is extremely
rapid, as is the equivalent for protons, $t_{ii}$.  These are the
timescales on which Maxwellian distributions are produced (e.g.,
Spitzer 1978).  But electron-ion \cb{temperature
  equilibration}\index{temperature equilibration} takes place on a
much longer timescale, so electrons and ions can have different
temperatures for times not short compared to the ages of young
supernova remnants.  (Once full temperature equilibration has been
attained, the gas will have the temperature given by the
Rankine-Hugoniot shock jump conditions, $kT_s = (3/16)\mu m_p v_s^2$,
where $\mu m_p$ is the mean mass per particle behind the shock.)
Observationally, electron temperatures determine the ionization state
of the gas, and strengths of lines, while ion temperatures can only be
inferred from line widths, if Doppler broadening due to bulk motions
can be removed.  Typical electron temperatures deduced in young SNRs
are several keV, though $kT_s$ can be 20 keV or higher (e.g., Rakowski
2005).

The emission produced by such ejecta is a combination of
bremsstrahlung continuum and line emission characteristic of the
ionization state of the gas.  As with electron-ion temperature
equilibration, ionization is not instantaneous, and SNR plasmas are
very frequently ionizing (that is, their ionization state lags behind
that of an equilibrium plasma at the same temperature; Itoh 1977).
Ionization is typically quite rapid up to helium-like states of common
elements, though lithium-like states of iron can be present as well.
So a typical SNR X-ray spectrum at CCD energy resolution ($E/\nabla E
\sim 20$) is dominated by blends of helium-like triplets ($K\alpha$
lines) of O, Ne, Mg, Si, S, Ar, and Ca.  (See the chapter ``X-ray
Emission Properties of Supernova Remnants'' in this volume.)  In SNe
Ia, L-shell transitions of Fe tend to produce a broad peak around 1
keV.  Grating or microcalorimeter energy resolution ($E/\nabla E
\gapprox 100$) is required to resolve these triplets.  Since ejecta
contain much higher proportions of heavier elements, those line
complexes are normally substantially stronger than the continuum, as
compared with solar-abundance plasma.

Temperatures in the keV range are reached for ejecta densities of
order 1 cm$^{-3}$.  If clumps are present, such as the ``FMK's''
(fast-moving knots) seen in Cas A, the densities can be very much
higher -- high enough that the shocks driven into clumps may have
speeds down to a few km s$^{-1}$.  Such a shock will be radiative, and
the resulting clump emission can be primarily observed in forbidden
lines of ions such as O$^+$ and O$^{+2}$.  
Post-shock densities in such clumps can be
$10^3$ cm$^{-3}$ or greater (Peimbert \& van den Bergh 1970).

If dust is formed in the cold unshocked ejecta, it can be radiatively
heated by the local UV photon field, or more likely, collisionally
heated once it passes through the reverse shock (e.g., Dwek \& Arendt
1992).  Dust in unshocked ejecta has been detected in a few cases;
since the temperatures are only a few tens of K, this radiation is in
the far-IR.  The {\sl Herschel} mission has found evidence for cold
dust in a few SNRs (e.g., SNR 1987A: Matsuura et al.~2011).  However,
in most cases, the infrared continuum one expects for ejecta dust
heated in the reverse shock is not present, and limits can be set on
the mass of dust produced in the SN and surviving the passage of the
reverse shock (e.g., Gomez et al.~2012).

{\bf Blast wave emission.}  The forward shock also
heats surrounding CSM or ISM to comparable temperatures, normally
somewhat higher than in the reverse shock.  Here too, the issues of
incomplete temperature or ionization equilibrium are important.
A typical pre-Sedov SNR will show a thermal X-ray spectrum which is
a complex mix of blast-wave and reverse-shock emission, with 
varying temperatures.  Decomposing that complex spectrum into
the constituent parts is a demanding task.

For a CC remnant, the blast wave normally encounters ionized material.
For a red supergiant (RSG) progenitor, the pre-explosion spectrum
includes relatively little ionizing radiation, but the shock breakout
of the RSG envelope will produce a UV flash that can ionize all the
surrounding material (normally the RSG wind) out to a distance that
can be many pc.  A compact progenitor that has lost most of its
envelope would not produce such a UV flash, but would have had a much
stronger pre-explosion ionizing flux (Reynolds et al.~2007).  

In contrast, all Type Ia SNRs less than a few thousand years old show
emission from nonradiative or \cb{Balmer-dominated
  shocks,}\index{Balmer-dominated shocks} resulting from partially
neutral upstream gas (Heng 2010).  Neutral H atoms do not feel the
magnetically mediated collisionless shock and remain at rest as the
shock sweeps over them.  On being suddenly immersed in $\sim 10^7$ K
plasma, most are immediately collisionally ionized, but a few are
first excited and can emit Balmer-series photons before being ionized.
Those photons show a Doppler broadening reflective of the (cold)
pre-shock distribution, resulting in a narrow line.  Some H atoms are
ionized by charge exchange, resulting in a fast neutral atom which can
also emit Balmer photons, but with a velocity distribution
characteristic of the downstream proton distribution, resulting in a
broad line.  The relative strengths, widths, and centroids of broad
and narrow components of lines from nonradiative shocks contain a
great deal of information about the upstream neutral fraction, the
shock geometry, and the up and downstream temperatures.  See Heng
(2010) for a review.  For young remnants, the presence of
nonradiative, Balmer-dominated shocks is strong evidence in favor of a
Type Ia origin.  (See the chapters ``Supernova/supernova remnant
connection'' and ``Supernova remnants as clues ot supernova
progenitors'' for more information on typing SNe from their remnants.)

SNR blast waves are virtually always easily detectable by their
synchrotron radio emission.  For the typical magnetic fields of tens
to hundreds of microGauss, this requires electron energies of order 1
-- 10 GeV; electrons radiating their peak synchrotron energy at
frequency $\nu$ have energies $E = 14.7(\nu_{\rm GHz}/B_{\mu{\rm
    G}})^{1/2}$ GeV.  For most remnants, ongoing electron acceleration
is required; the interstellar cosmic-ray electron energy spectrum is
considerably flatter than what is seen in SNRs, and for young remnants
with adiabatic shock waves in particular, the limited shock
compression means that compressing ambient magnetic field and
electrons cannot produce the observed radio surface brightnesses
typical of young remnants (Reynolds 2008).  The radio spectra are well
described by power-laws, $S_\nu \propto \nu^{-\alpha}$ with $\alpha
\sim 0.4 - 0.7$ for most objects.  Young SNRs tend to have steeper
radio spectra; Cas A has $\alpha = 0.77$.  See Green (2014) for an
extensive compilation of observations of 294 Galactic SNRs.

The theory of diffusive shock acceleration (DSA) is generally thought
to be responsible for the relativistic particle populations we infer
in SNRs.  The review by Blandford and Eichler (1987) is still an
excellent introduction. If nonthermal particles make up a small
fraction of the postshock energy density (the ``test-particle
limit'',) DSA predicts a particle spectrum $N(E) \propto E^{-s}$ with
$s = 2$ in a strong shock (Mach number ${\cal{M}} \gg 1$) with
compression ratio $r = 4$. The synchrotron spectrum from such a
population of electrons is a power-law with $\alpha = (s-1)/2 = 0.5.$

However, if SNRs are the source of Galactic cosmic rays, the total
Galactic content of cosmic ray energy requires that of order 10\% of
SNR energy be put into fast particles -- too large for the
test-particle limit.  In this case, the fast particles can modify the
shock structure as they diffuse ahead of the shock, producing a
gradual rather than sudden change in flow velocity, until a ``viscous
subshock'' with a compression ratio of 2 to 3 and thickness of a few
ion mean free paths heats the gas.  Then if (as is likely) particle
mean free paths increase with energy, more energetic particles diffuse
further ahead of the shock before being scattered back, sampling a
larger compression ratio and forming a locally harder spectrum.  That
is, the energy distribution of accelerated particles becomes concave
up, steeper than the test-particle limit at low energies and flatter
at high energies.  This effect could produce the steeper radio spectra
of young SNRs.  However, the large numbers of older SNRs with $\alpha
< 0.5$ are still difficult to explain.  Contamination of the spectrum
with optically thin thermal bremsstrahlung ($\alpha = 0.1$) at higher
energies could be responsible in some cases.

\begin{figure}
%\centerline{\includegraphics[scale=0.6]{modshockfig.eps}}
\centerline{\includegraphics[scale=0.6]{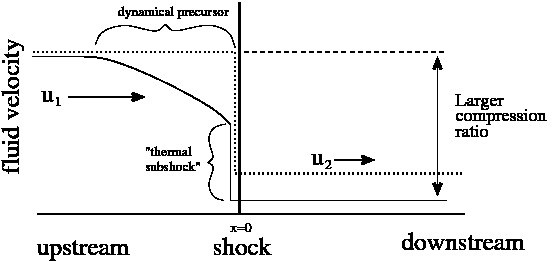}}
\caption{Schematic of a cosmic-ray modified shock.  In the shock
frame, material enters from the left and is gradually decelerated
by cosmic rays diffusing upstream in a ``dynamical precursor''
until a sharp drop in speed at the thermal subshock.  The
overall shock compression ratio can be considerably larger than
in an unmodified shock.}
\label{fig:modshock}
\end{figure}

The maximum energy to which particles can be accelerated depends on
the shock age, magnetic field, and other properties (Reynolds 1998).
The time $t_{\rm acc}(E)$ to reach an extremely relativistic energy
$E$, for both electrons and protons, depends on the diffusion
coefficient $\kappa(E)$ and shock speed $v_s$, where $\kappa =
\lambda_{\rm mfp}c/3$.  The mean free path $\lambda_{\rm mfp}$ is
often assumed proportional to the particle gyroradius $r_g$,
$\lambda_{\rm mfp} = \eta r_g$, (``Bohm-like'' diffusion, with $\eta =
1$ giving the ``Bohm limit.'')  In this case, $t_{\rm acc} \sim
\kappa(E)/v_s^2$, so fast shocks can produce much higher energies.
Since $r_g = E/eB$ for relativistic particles (cgs units; $e$ is the
electronic charge), high magnetic fields also produce more rapid
acceleration.  For ions, radiative losses are insignificant; the
limitation is basically the shock age, $t = t_{\rm acc}$.  For
electrons, radiative losses due to synchrotron radiation or
inverse-Compton upscattering of any local photon fields can limit the
maximum energy much more severely.  For synchrotron losses, $E_{\rm
  max} \propto B^{-1/2}$.  However, since an electron of energy $E$
radiates its peak synchrotron power at a frequency $\nu \propto E^2
B$, the peak frequency $\nu_{\rm max}$ radiated by a distribution of
shock-accelerated electrons limited by losses is independent of the
magnetic-field strength $B$.

For young remnants with shock velocities of order 1000 km s$^{-1}$ or
greater, $h\nu_{\rm max}$ can easily exceed 1 keV, so synchrotron
radiation is produced all the way from radio to X-ray energies, with
electron energies reaching 10 TeV or more (Reynolds \& Chevalier
1981).  Thermal emission at optical and infrared wavelengths normally
swamps this contribution, although near-IR observations of Cas A have
identified a synchrotron contribution.  However, in X-rays, a handful
of Galactic remnants are dominated by synchrotron emission (including
most notably the youngest Galactic SNR, G1.9+0.3), while all
historical shell SNRs show local regions dominated by synchrotron --
usually, but not always, in ``thin rims'' at the location of the blast
wave.  If the rims are thin because electrons rapidly lose energy as
they advect downstream, magnetic field values of 100 $\mu$G or higher
are inferred (Parizot et al.~2006).

These particle energies can result in significant photon emission
above the X-ray region, at GeV and TeV energies.  See Reynolds (2008)
for a review.  Depending on the energy density of the local radiation
field, inverse-Compton scattering by the same relativistic electrons
can make substantial contributions, with a minimum set by upscattering
of cosmic microwave background photons (``ICCMB'').  In addition,
mildly relativistic electrons (the same that produce radio
synchrotron) can produce a nonthermal bremsstrahlung contribution.
Protons and other ions ought to reach energies at least as high as
electrons in DSA.  Relativistic ions do not radiate significantly, but
can produce pions through inelastic collisions with thermal gas, once
they reach the energy threshold of about 70 MeV.  The neutral pions
decay to gamma-rays which can be detected.  There are currently a
dozen or more SNRs with detected gamma-ray emission in the Fermi LAT
band (GeV) or by ground-based air-{\v C}erenkov detectors (TeV).  The
question of whether the gamma-ray emission from these objects is due
to leptons or hadrons is actively discussed.  Hadronic domination
requires substantial thermal target densities.  Escaping particles
ahead of the SNR blast wave may impinge on dense clouds to produce
emission in some cases.

The evolution of the synchrotron radio emission from a SNR is
straightforward to estimate, for various possible assumptions about
the efficiency of shock acceleration and magnetic-field amplification.
The synchrotron volume emissivity due to a power-law energy
distribution of electrons $N(E) = KE^{-s}$ cm$^{-3}$ erg$^{-1}$ can be
conveniently written $j_\nu = c_j (\alpha) K B^{1 + \alpha}
\nu^{-\alpha}$ (e.g., Pacholczyk 1970).  Then the flux density from
a spherical remnant at distance $d$ is given by
\begin{equation}
S_\nu = (4\pi j_\nu)\left(  R_s^3 \phi \over 3d^2 \right)
\end{equation}
where $\phi$ is the volume filling factor of emitting material ($\phi
\sim 0.25$ for a Sedov remnant with shock compression ratio 4).  If
the shock puts constant fractions $\epsilon_B$ and $\epsilon_e$ of the
post-shock pressure $\rho v_s^2$ into magnetic-field energy and
electron energy, respectively, and if the upper and lower bounds on
the electron distribution $E_h$ and $E_l$ do not change, we have
\begin{equation}
S_\nu \propto \epsilon_B^{(1 + \alpha)/2}\epsilon_e 
    R_s^3 \rho^{(3 + \alpha)/2} v_s^{3 + \alpha}
\propto t^{m(6 + \alpha) - (3  + \alpha)}
\end{equation}
since if $R_s \propto t^m$, then $v_s \propto t^{m - 1}$.  This
assumes constant ambient density.  For a remnant with a typical value
of $\alpha = 0.5$, $S_\nu \propto t^{6.5m - 3.5} = t^{-0.9}$ for a
Sedov remnant ($m = 0.4$), and {\em rises} with time for an
ejecta-driven remnant with $m > 0.54$ (although the simple assumptions
made here may not hold for such early times).  For a remnant
encountering a steady wind with $\rho \propto r^{-2}$, we have
$S_\nu \propto t^{4.5m - 3.5}$ (still with $\alpha = 0.5$) which
virtually never increases.  

Other assumptions are possible.  If the magnetic field is not
amplified but is simply compressed from a uniform value upstream, and
the ambient density is uniform, then $S_\nu \propto t^{5m-2}$ which is
constant in the Sedov phase.  The shock may accelerate all electrons
with energies above some threshhold which varies with shock velocity.
Or efficiencies may change with time.  It is not known at present
which of these assumptions is correct.  At this time, only the
Galaxy's youngest SNR, G1.9+0.3, is brightening with time at radio
frequencies (Murphy et al.~2008), while the next three youngest, Cas
A, Tycho, and Kepler, are all fading, at rates between 0.2 and 0.7 \%
yr$^{-1}$ (Vinyaikin 2014; Stankevich et al.~2003).  (An exception is
the SN(R) 1987A in the LMC, which is brightening at both radio and
X-ray wavelengths (Zanardo et al.~2010; Helder et al.~2013).

For reasons still not completely clear, reverse shocks are not obvious
particle accelerators.  The theory of DSA has been very successful in
interpreting forward-shock nonthermal emission, but there may be
circumstances in which other processes, such as stochastic or
turbulent acceleration, play a role.  The nonthermal X-ray emission in
Cas A above 20 keV (Grefenstette et al.~2015) does not appear to be
associated with either the forward or reverse shocks, and is still not
fully understood.

Infrared continuum emission can be produced by any dust present either
in the ejecta or the ISM/CSM.  Many remnants show emission in mid-IR
bands such as {\sl Spitzer}'s 24 $\mu$m band which is morphologically
well-correlated with radio emission, and is consistent with being
produced by collisional heating of surrounding dust by the
forward-shocked plasma (Borkowski et al.~2006, Williams et al.~2006).
This heating process depends strongly on the plasma density and much
less strongly on the plasma temperature, so IR spectra, or even
two-point color temperatures, can provide powerful diagnostics of SNR
densities.  The results can be surprising; the symmetric circular
outline of Tycho's SNR masks density variations of a factor of 6 or
greater (Williams et al.~2013).

{\bf Later stages.}  Once the reverse shock has disappeared and all
ejecta have been reheated, a spherical remnant can be well described
by Sedov profiles of density and temperature.  Typically by this
stage, gas in the outermost 10\% or so of the radius (most of the
material) is dense enough that electron and ion temperatures have come
into equilibration, and ionization equilibrium has been reached as
well (that is, the ionization timescale $\tau \equiv \int n_e\,dt
\gapprox 3 \times 12$ cm$^{-3}$ s).  While this considerably
simplifies X-ray spectral analysis, the plasma temperature still
varies widely, rising from its immediate postshock value toward the
interior.  For the fast shocks of young remnants, single-temperature
plane shocks may provide adequate descriptions in restricted
bandpasses. 

Structure in the surrounding medium, either modified CSM or
pre-existing inhomogeneities, can result in non-monotonic evolution of
the blast wave speed.  If the shock breaks out of a denser region into
a less dense one, rapid adiabatic cooling can leave the shocked plasma
in an overionized state.  Spectral diagnostics of this state,
including radiative recombination continuum and line ratios
inconsistent with temperatures derived from X-ray continua, have been
seen in a few SNRs, of which one of the first was the middle-aged
remnant W49B (Ozawa et al.~2009).

The primary change in emission properties of an SNR at late stages is
the appearance of bright optical emission as the shocks become
radiative and roughly isothermal.  That is, an initial jump in density
of a factor of 4 at the shock is followed downstream by a much larger
density increase in the ``cooling layer,'' where much of the shock
energy is radiated.  This layer can be identified by bright emission
from such species as O$^{+2}$.  Compressions can reach factors of 100
or more, so regions with densities of $10^3$ cm$^{-3}$ and higher now
dominate the remnant spectrum. Shock speeds are now in the range of
100 km s$^{-1}$ or lower; optical diagnostics of various line ratios
are available to characterize the temperature and density of such
regions.  Radiative shocks are complex and heterogeneous, typically
involving a superposition of shock speeds, but models do a fairly good
job of accounting for ratios of line strengths of many species that
can be observed (Cox \& Raymond 1985; Innes et al.~1987).

The high compressions mean that radio emission can be quite bright as
well, as even without additional electron acceleration, energy
densities of ambient cosmic-ray electrons and magnetic field can be
increased by large factors (van der Laan 1962).  Gas densities from
optical diagnostics heavily favor very dense regions that occupy
relatively little volume, so are probably not typical of most of the
radio-emitting volume.  With shock velocities of 100 km s$^{-1}$ or
lower, ongoing particle acceleration is probably weak at best.  Some
tendency of the very oldest SNRs to have the flattest radio spectra
may simply reflect the greater importance of thermal radio
contamination (Oni\'c 2013).

Most extragalactic SNRs are found with methods that favor large
optically bright radiative remnants.  (See ``Surveys of Supernova
Remnants and Detection Techniques'' in this volume.) These objects
tell us as much about the homogeneity and character of surrounding ISM
as about the nature of the supernova or its progenitor system.
However, since most remnants spend most of their detectable lifetimes
in these stages, statistics of SNRs do not suffer terribly.  Most
important in the analysis of such statistics is the range of ambient
densities into which a population of SNRs may be evolving.  Most of
the scatter in distributions of SNRs in plots such as the \cb{surface
  brightness-diameter (``$\Sigma-D$'')
  relation}\index{surfac-brightness-diameter (``$Sigma-D$'') relation}
is probably caused by variations in upstream density, making these
relations unreliable at best for inferring basic information about SNR
evolution.

Subsequent chapters will examine in more detail these various issues.

\section{Conclusions}
\label{sec:4}

The traditional outline of SNR evolution from free expansion to Sedov
evolution to radiative snowplow provides only a crude description of a
continuous development in which ejecta immediately begin interacting
with CSM, with the rapid formation of a reverse shock.  Deceleration
of the outer blast wave begins in a few years, so there is no real
free expansion (the expansion parameter $m$ is less than 1 almost from
the beginning, and smoothly evolves toward its Sedov value of 0.4).
The details of this evolution depend on the ejecta density structure.
The reverse shock eventually moves to the remnant center and reheats
all ejecta, though this may not occur until many times the ejected
mass has been swept up.  The blast wave is a strong source of thermal
X-rays and nonthermal radio emission, and for young remnants, also
nonthermal X-ray and gamma-ray emission.  The reverse shock produces
strong thermal X-ray emission as well.  Until ionization equilibrium
is reached, X-ray emissivities from both shocks can be much higher
than for equilibrium plasmas.  Since cooling times are a strong
function of density, for an older remnant encountering inhomogeneous
ISM, some regions (``clumps'' or ``clouds'') will become radiative
sooner than others, and optical emission will be dominated by small
regions of very high, atypical, densities.  Thus the optical
luminosity of an older SNR is not a good indicator of its global
evolutionary state.  Thermal X-ray emission from heated ISM and ejecta
can remain detectable even after much of the shock is radiative.
Improving our understanding of SNRs, of the supernovae that produce
them, and of the CSM and ISM with which they interact, requires more
realistic descriptions of both evolution and radiation.

\section{Acknowledgments}
\label{sec:5}

I am grateful for discussions with many colleagues over many years,
among whom Roger Chevalier, Kazimierz Borkowski, John Blondin, and
Roger Blandford are particularly prominent.  I am pleased to
acknowledge support from the National Science Foundation and National
Aeronautics and Space Administration for research support over the
last 30 years.

\section{Cross-references}
\label{sec:6}

All the succeeding chapters in this Section elaborate various topics I
have touched on here.  The Dynamical Evolution section contains
references to Chapter 7, Pulsar Wind Nebulae, Chapter 8, Supernova
Remnant from SN 1987A, Chapter 9, Supernova/Supernova Remnant
Connection, and Chapter 10, Supernova Remnants as Clues to Supernova
Progenitors.  The Radiative Evolution section is elaborated in
Chapters 2 -- 6: Surveys, Radio Emission, X-ray Emission, UV and
Optical Emission, and Infrared Emission.

%The extensive Galactic census of SNRs by Green
%(http://www.mrao.cam.ac.uk/surveys/snrs/) lists spectral indices

%such
%as [Fe II] 1.26, 1.64, and 26 $\mu$m, [Ar II] 6.99 $\mu$m, 

\input{referenc}
\end{document}

%% file: referenc.tex
%%%%%%%%%%%%%%%%%%%%%%%% referenc.tex %%%%%%%%%%%%%%%%%%%%%%%%%%%%%%
% sample references
% %
% Use this file as a template for your own input.
%
%%%%%%%%%%%%%%%%%%%%%%%% Springer-Verlag %%%%%%%%%%%%%%%%%%%%%%%%%%
%
% BibTeX users please use
 \bibliographystyle{}
 \bibliography{}